# Low mode volume slotted photonic crystal single nanobeam cavity

Judson D. Ryckman and S. M. Weiss

*Department of Electrical Engineering and Computer Science, Vanderbilt University, Nashville, TN 37235, USA*

We present and experimentally demonstrate a slotted photonic crystal single nanobeam cavity in silicon. The slot geometry is exploited to achieve ultra-small effective mode volumes, $\sim 0.025(\lambda/n)^3$, more than an order of magnitude smaller than traditional nanobeam cavities. A continuous slot and a tapered photonic crystal design are implemented to achieve experimental quality factors near $10^4$. This device structure offers a unique platform for achieving enhanced light-matter interactions and could be used to benefit a variety of applications including non-linear optics, cavity quantum electro-dynamics, sensing, optical modulation, and nano-optomechanics.

In recent years, photonic crystal (PhC) nanobeam cavities have emerged as an advantageous platform for enhancing light-matter interaction, owing to their compact size, ultra-high Q-factors, and low mode volumes $\sim (\lambda/n)^3$.[1-3] Achieving small mode volumes is often pursued for applications related to nonlinear optics[4,5] and cavity quantum electrodynamics (QED),[6] where increasing the figure of merit Q/V is critical. Small mode volumes are also important for reducing the footprint and energy requirements of optical modulators[7-9] and lasers,[10,11] reducing the molecular detection limit of nanosensors[12,13] and increasing optical forces for optical trapping[14] and opto-mechanical coupling.[15,16] A primary strategy for reducing the effective mode volume, as defined in Ref. 17, is to introduce a dielectric discontinuity, or slot, which serves to simultaneously enhance the maximum field intensity and localize the maximum field to a region of lower refractive index. This approach has been demonstrated in 2D photonic crystal (PhC) slabs[18,19] by introducing a slot into the center of a W1 waveguide. Heterostructure cavities based on the slotted 2D PhC platform have recently been implemented for enhanced optical modulation,[20] sensing,[21] and optomechanical coupling.[15] Compared to 2D PhC cavities, PhC nanobeam cavities offer a simpler geometry, reduced device footprint, and lower total mass, which make them an attractive alternative to 2D PhC cavities for many of the aforementioned applications. With recent improvements to the understanding and design of ultra high-Q 1D PhCs,[22,23] ultra-small mode volume slotted nanobeam cavities, which maintain high Q-factors, have recently been theorized but not yet demonstrated.[24,25]

Coupling two nanobeams together has been shown to produce bonded (even) and anti-bonded (odd) supermodes, the former of which exhibits strong field confinement in the slot-like gap between nanobeams.[1,26,27] While cavities based on the bonded (even) modes of double nanobeams feature localized field enhancements and strong gradient optical forces, the smallest achievable effective mode volumes, occuring when the slot between nanobeams approaches zero width, are on the order of $0.1(\lambda/n)^3$,[26] only a modest improvement over the $V_{eff} \sim 0.3–1(\lambda/n)^3$ values for single nanobeam cavities.[1, 3] Although the field enhancement in the slot between coupled nanobeams increases with decreasing slot width, the individual nanobeams necessarily contain a sizeable fraction of the optical power because they are evanescently coupled. Therefore, a slotted *single* nanobeam cavity should be able to achieve a smaller mode volume than double nanobeam cavities.

In this letter we present the design and experimental demonstration of a slotted ultra-small mode volume single nanobeam cavity in silicon, achieving a mode volume $V = 0.025(\lambda/n)^3$ and experimental $Q = 7.42 \times 10^3$. The concept of introducing a slot into a 1D PhC cavity was originally proposed by Robinson *et al.*,[17] where it was shown that the effective mode volume can theoretically be reduced by a factor of $(n_{hi}/n_{lo})^5$ for an infinitesimal slot width. Unfortunately, introducing a finite slot, without any further design tailoring, can reduce the theoretical Q-factor by over three orders of magnitude, thereby limiting the Q-factor to be on the order of $\sim 10^2$.[28] Such low Q-factors are almost entirely a result of increased radiation losses, which occur because the introduction of a finite slot causes a large perturbation to the optical mode and introduces spatial frequencies inside the light cone.[29, 30] In order to reduce leakage to radiation modes, and improve the radiation quality factor $Q_r$, we have incorporated two important design choices into the present work: (1) the use of a common five-hole taper design[1, 2, 22] and (2) the use of a continuous slot which is unterminated in the principle direction of the photonic band gap (PBG).[25] Both of these design choices present a gentler way of breaking the translational symmetry of the PhC mirror and serve to reduce the typically large mismatch between the effective index of the defect and evanescent Bloch mode in the mirror. Moreover, the continuous slot geometry promotes efficient input/output coupling to slot waveguides.

Fig. 1(a) shows the dispersion relation for the slotted photonic crystal waveguide structure used in this work. This waveguide structure, which we refer to as the *pinch waveguide*,[25] exhibits guided modes that lie outside the air (n = 1) light line and shows a sizeable $\Delta\omega \approx 0.063$ mode-gap. The dispersion relation is obtained from a 3D band structure calculation for quasi-TE polarization and a periodic unit cell of length *a*, with waveguide width *w = 1.091a*, hole radius *r = 0.335a*, slot width *ws = 0.093a*, and waveguide thickness *h = 0.379a* (corresponding to the mirror dimensions of the device shown and tested in Figs. 2&3). A refractive index $n_{Si}$ = 3.46 is used for silicon in all simulations. These dimensions were selected to balance three primary constraints: (1) the predefined SOI device layer thickness ~212nm, (2) a minimum reproducible slot width ~50nm, and (3) the ability to place the dielectric band edge near ~1550nm. The connected nature of the basic holey waveguide structure promotes the largest photonic band gaps for quasi-TE polarization.[31] Moreover, quasi-TE polarization, where the dominant electric field component, $E_y$, is oriented in the plane, gives rise to a strong slot-type field enhancement.[32] Fig. 1(b) reveals the field and power distribution for a mode on the *dielectric band* at the edge of the first Brillouin zone. The dielectric band is characterized by strong field confinement and localized field enhancement within the slot, located in the "pinch" region between neighboring air holes. These favorable field characteristics in the pinch region are conducive to low mode volume cavities. As reported previously, the pinch waveguide is capable of providing a normalized $|E_{max}|^2$ enhancement up to ~4-5 times greater than that of a traditional slot waveguide of the same nominal dimensions.[25] Thus, according to the definition of the effective mode volume,[17] with $V_{eff} \propto 1/|E_{max}|^2$, the mode volume of a pinch waveguide can be up to ~4-5 times smaller than that of a comparable standard slot waveguide.

Our devices were fabricated from silicon-on-insulator (SOI) wafers with a ~212nm thick Si device layer and ~1μm thick buried oxide. Electron beam lithography (JEOL JBX-9300-100kV) was performed using ZEP 520A resist spun at 6,000rpm (~300nm thick). Reactive ion etching of the exposed Si regions was performed with $C_4F_8$, $SF_6$, and Ar process gases. Photolithography was then carried out using S1813 photoresist to facilitate selective undercutting of the nanobeam while not affecting the input/output bus waveguides and grating couplers. A standard 10:1 buffered oxide etchant (BOE) was applied in two ten-minute intervals with deionized (DI) water rinsing and nitrogen drying performed in between and after the BOE steps. This approach was sufficient to release the ~30μm length nanobeams without the need for vapor etching or critical point drying. Finally, samples were soaked and agitated in acetone for 5min to remove the photoresist, and immediately rinsed in DI water and dried under nitrogen flow.

Our slotted single nanobeam cavity, shown in Fig. 2, is formed utilizing a pair of Bragg mirrors, consisting of N = 9 mirror segments, with lattice constant $a$ = 559nm. This choice of $a$ places the dielectric band edge near 1550nm. Toward the center of the cavity five-hole tapers are used, wherein the hole spacing is linearly tapered from 559nm to a defect spacing of 450nm between the central most air-holes. At the same time, the hole radii are tapered from approximately 187.5nm to 150nm. On either side of the Bragg mirrors, input/output tapers are used to promote more efficient coupling to slot waveguides.[22] These tapers utilize the same linear reduction in hole spacing as the central cavity tapers, but with a linear reduction in hole size from 187.5nm radius to 110nm to better match the effective index of the slot waveguide. Across the entire device length, a slot width of 52nm and waveguide width of 610nm are used.

Lastly, the slot waveguides are efficiently coupled to input/output bus ridge waveguides through compact mode converters[33] that are routed to input/output grating couplers oriented at 90° to facilitate cross-polarized optical characterization.

As shown in Fig. 2(b) the slotted PhC single nanobeam cavity structure supports a fundamental mode with a high Q-factor ~$3.5\times10^4$ and ultra-small mode volume ~$0.025(\lambda/n)^3$, determined from 3D finite difference time domain (FDTD) simulation. In our simulations, we found that the device Q-factor was very sensitive to the exact defect spacing and taper geometry, similar to reports for non-slotted nanobeam cavities.[1, 3] The Q-factor reported here is more than one order of magnitude higher than a recently proposed design with the same slot width.[24] By using an even smaller slot width, a simultaneous increase in Q-factor and reduction in mode volume could be achieved. However, due to fabrication feature-size limitations, it may be desirable to realize higher Q-factors by instead employing a more gradual tapering strategy or a deterministic design approach.[3, 23, 34] We anticipate that the theoretical cavity Q-factor of a slotted PhC single nanobeam cavity can be increased to match the recently reported ultrahigh >$10^6$ Q-factors reported for other 1D PhC cavities. However, any substantial increase in Q-factor will likely come at the cost of an increased modal volume.

Slotted PhC single nanobeam cavities were measured using a cross-polarized optical configuration, which was found to be a rapid and straightforward method that enabled the measurement of many devices within a compact area, thus eliminating the need for ~mm length input/output waveguides and precisely aligned lensed fibers. Light from a tunable laser (Santec TSL-510) is sent through a beam splitter and focused onto the input grating coupler, Fig. 2(a), through a long working distance 20x objective. From infrared imaging, the spot size was estimated to be ~30μm in diameter. With the same objective and beamsplitter, light from the sample surface is simultaneously collected, passed through a linear polarizer oriented 90° relative to the input polarization, and analyzed by an InGaAs photodetector camera.

Fig. 3(a) shows the optical transmission of the slotted photonic crystal single nanobeam cavity normalized by the peak optical transmission near the band edge. Similar to other recently demonstrated nanobeam cavities, simulations suggest approximately unity optical transmission near the band edge.[3, 35] The measurement shown in Fig. 3(a) reveals the dielectric band edge at ~1548nm, in close agreement with simulations (Fig. 1(a)), and the presence of a single high Q-factor resonance, at 1513.9nm, with a peak normalized optical transmission of ~0.12. This relatively low resonance transmission level, which can be described by $T = (Q_{meas} / Q_{wvg})^2$,[36] suggests that the cavity Q-factor is not limited by the waveguide Q-factor, $Q_{wvg}$, but rather the radiation Q-factor, $Q_{rad}$. Note that $1/Q_{meas} = 1/Q_{rad} + 1/Q_{wvg}$. In other words, the lifetime for a photon to be scattered out of the cavity is shorter than the lifetime for a photon to decay into the waveguide. This is further apparent when examining the IR camera image of the cavity on resonance [inset; Fig. 3(a)], as we observe a larger intensity for vertical radiation coming from the nanobeam than we do for light coming from the output grating coupler. As shown in Fig. 3(b), our optical interrogation technique also enables us to measure vertical radiation spectra for light radiating out of the cavity. Whereas other configurations might isolate the collected signal using a pinhole and photodiode,[37] our scheme relies on the position sensitive detection capability of the photodetector camera, similar to what has been reported for measurements on multiple ring resonator systems.[38] The vertical radiation spectra similarly reveals the single high Q-factor resonance at 1513.9nm, but with an even higher signal-to-noise ratio, confirming that both the measured Q-factor and ~0.12 transmission level are limited by the radiation losses. Lorentzian fitting of both resonance measurements reveals a measured quality factor, $Q_{meas} \sim 7.42 \times 10^3$. Combined with the ~0.12 on-resonance optical transmission, we estimate an intrinsic radiation Q-factor, $Q_{rad} \sim 1.14 \times 10^4$. Although this quality factor is smaller than the record high Q-factors reported for non-slotted 1D PhC cavities,[1, 39] our experimentally achieved $\sim 10^4$ Q-factor is already sufficient to enable applications ranging from optical modulation to sensing, where the ultra-small $\sim 0.025(\lambda/n)^3$ mode volume should offer substantial improvements to device sensitivity.

The slotted PhC single nanobeam cavity also offers a unique geometry for nano-optomechanics applications. Notably, our nanobeam cavity exhibits stronger field confinement and a lower total mass (<10pg) than previously demonstrated nanobeam cavity configurations (typically ~10-50pg)[16, 37]. Because the slot mechanically separates each half of the nanobeam, we anticipate that optical forces could be exploited to manipulate the slot width and tune the cavity, or achieve strong opto-mechanical coupling, similar to what has been reported for dual nanobeam implementations.[26, 27, 37, 40] Given the lower mode volume of the slotted single nanobeam cavity and its increased sensitivity to the exact slot dimensions, we anticipate that the optical response will be much more sensitive to mechanical displacements. This should enable enhanced tunability for electrically programmable nanobeam cavities[41] and all-optically reconfigurable optomechanical filters.[40]

In summary, we have reported the demonstration of a slotted photonic crystal single nanobeam cavity in silicon. The slot geometry is exploited to provide strong confinement and localized field enhancements, resulting in an ultra-small mode volume $\sim 0.025(\lambda/n)^3$. This represents the smallest mode volume ever achieved in an all-dielectric 1D PhC cavity while still maintaining a Q-factor $>10^3$. The high Q-factor is achieved by combining a common five-hole taper design with the continuously slotted pinch waveguide platform, and we expect further design tailoring can be used to achieve ultra-high Q-factors $>10^5$. The slotted PhC cavity platform could be used to benefit a variety of applications including non-linear optics, cavity QED, sensing, optical modulation, and nano-optomechanics.

Acknowledgement:

This work was supported in part by the Air Force Office of Scientific Research under Grant FA9550-10-1-0366. The authors further acknowledge support from ARO DURIP equipment grant W911-NF-10-1-0319. Portions of this work were performed at the Vanderbilt Institute of Nanoscale Science and Engineering, using facilities renovated under NSF ARI-R2 DMR-0963361, and the Center for Nanophase Materials Sciences, which is sponsored at Oak Ridge National Laboratory by the Division of Scientific User Facilities, U.S. Department of Energy. J.D.R. acknowledges support from a NSF Graduate Research Fellowship.

Reference:

1. P. B. Deotare, M. W. McCutcheon, I. W. Frank, M. Khan and M. Loncar, Appl. Phys. Lett. **94**, 121106 (2009).
2. S. Gary, E. Bryan, P. Jan, A. M. Marie, S. Tomas, H. James, E. H. Eugene and J. Vuckovic, Appl. Phys. Lett. **99**, 071105 (2011).
3. Q. M. Quan and M. Loncar, Opt. Express **19**, 18529 (2011).
4. I. B. Burgess, Y. N. Zhang, M. W. McCutcheon, A. W. Rodriguez, J. Bravo-Abad, S. G. Johnson and M. Loncar, Opt. Express **17**, 20099 (2009).
5. T. Tanabe, M. Notomi, S. Mitsugi, A. Shinya and E. Kuramochi, Appl. Phys. Lett. **87**, 151112 (2005).
6. R. Ohta, Y. Ota, M. Nomura, N. Kumagai, S. Ishida, S. Iwamoto and Y. Arakawa, Appl. Phys. Lett. **98**, 173104 (2011).
7. S. Manipatruni, K. Preston, L. Chen and M. Lipson, Opt. Express **18**, 18235 (2010).
8. M. Belotti, M. Galli, D. Gerace, L. C. Andreani, G. Guizzetti, A. R. M. Zain, N. P. Johnson, M. Sorel and R. M. De la Rue, Opt. Express **18**, 1450 (2010).
9. J. D. Ryckman, V. Diez-Blanco, J. Nag, R. E. Marvel, B. K. Choi, R. F. Haglund and S. M. Weiss, Opt. Express **20**, 13215 (2012).
10. S. Matsuo, A. Shinya, T. Kakitsuka, K. Nozaki, T. Segawa, T. Sato, Y. Kawaguchi and M. Notomi, Nat. Photon. **4**, 648 (2010).
11. S. Kita, S. Hachuda, K. Nozaki and T. Babab, Appl. Phys. Lett. **97**, 161108 (2010).
12. S. Mandal and D. Erickson, Opt. Express **16**, 1623 (2008).
13. J. N. Anker, W. P. Hall, O. Lyandres, N. C. Shah, J. Zhao and R. P. Van Duyne, Nat. Mater. **7**, 442 (2008).
14. S. Mandal, X. Serey and D. Erickson, Nano Lett. **10**, 99 (2010).

15. H. S.-N. Amir, P. M. A. Thiago, W. Martin and P. Oskar, Appl. Phys. Lett. **97**, 181106 (2010).
16. M. Eichenfield, R. Camacho, J. Chan, K. Vahala and O. Painter, Nature **459**, 550 (2009).
17. J. T. Robinson, C. Manolatou, L. Chen and M. Lipson, Phys. Rev. Lett. **95**, 143901 (2005).
18. T. Yamamoto, M. Notomi, H. Taniyama, E. Kuramochi, Y. Yoshikawa, Y. Torii and T. Kuga, Opt. Express **16**, 13809 (2008).
19. J. Gao, J. F. McMillan, M. C. Wu, J. J. Zheng, S. Assefa and C. W. Wong, Appl. Phys. Lett. **96**, 051123 (2010).
20. J. H. Wulbern, J. Hampe, A. Petrov, M. Eich, J. D. Luo, A. K. Y. Jen, A. Di Falco, T. F. Krauss and J. Bruns, Appl. Phys. Lett. **94** (2009).
21. A. Di Falco, L. O'Faolain and T. F. Krauss, Appl. Phys. Lett. **94**, 063503 (2009).
22. A. R. M. Zain, N. P. Johnson, M. Sorel and R. M. De la Rue, Opt. Express **16**, 12084 (2008).
23. M. Notomi, E. Kuramochi and H. Taniyama, Opt. Express **16**, 11095 (2008).
24. P. Yu, B. A. Qi, X. Q. Jiang, M. H. Wang and J. Y. Yang, Opt. Lett. **36**, 1314 (2011).
25. J. D. Ryckman and S. M. Weiss, IEEE Photonics J. **3**, 986 (2011).
26. J. Chan, M. Eichenfield, R. Camacho and O. Painter, Opt. Express **17**, 3802 (2009).
27. F. Kevin, L. Loic, C. Benoit, P. Emmanuel, P. David, F. Frederique de and H. Emmanuel, Appl. Phys. Lett. **94**, 251111 (2009).
28. A. Gondarenko and M. Lipson, Opt. Express **16**, 17689 (2008).
29. J. Vuckovic, M. Loncar, H. Mabuchi and A. Scherer, IEEE J. Quantum Elect. **38**, 850 (2002).
30. Y. Akahane, T. Asano, B. S. Song and S. Noda, Nature **425**, 944 (2003).
31. S. H. Fan, J. N. Winn, A. Devenyi, J. C. Chen, R. D. Meade and J. D. Joannopoulos, J. Opt. Soc. Am. B **12**, 1267 (1995).
32. V. R. Almeida, Q. F. Xu, C. A. Barrios and M. Lipson, Opt. Lett. **29**, 1209 (2004).
33. Z. C. Wang, N. Zhu, Y. B. Tang, L. Wosinski, D. X. Dai and S. L. He, Opt. Lett. **34**, 1498 (2009).
34. B. Desiatov, I. Goykhman and U. Levy, Appl. Phys. Lett. **100**, 041112 (2012).
35. Q. M. Quan, P. B. Deotare and M. Loncar, Appl. Phys. Lett. **96**, 203102 (2010).
36. S. G. Johnson, A. Mekis, S. H. Fan and J. D. Joannopoulos, Comput. Sci. Eng. **3**, 38 (2001).
37. Y. Y. Gong, A. Rundquist, A. Majumdar and J. Vuckovic, Opt. Express **19**, 1429 (2011).
38. M. L. Cooper, G. Gupta, J. S. Park, M. A. Schneider, I. B. Divliansky and S. Mookherjea, Opt. Lett. **35**, 784 (2010).
39. E. Kuramochi, H. Taniyama, T. Tanabe, K. Kawasaki, Y. G. Roh and M. Notomi, Opt. Express **18**, 15859 (2010).
40. P. B. Deotare, I. Bulu, I. W. Frank, Q. Quan, Y. Zhang, R. Ilic and M. Loncar, Nat. Commun. **3**, 846 (2012).
41. I. W. Frank, P. B. Deotare, M. W. McCutcheon and M. Loncar, Opt. Express **18**, 8705 (2010).

Figure captions:

FIG. 1 (a) Dispersion relation for the slotted PhC pinch waveguide, shown in the inset, with $w = 1.091a$, $r = 0.335a$, $ws = 0.093a$, and waveguide thickness $h = 0.379a$. (b) Field and power distributions for a single unit cell of the pinch waveguide structure calculated at the dielectric band edge.

FIG. 2 (a) SEM images of a slotted photonic crystal single nanobeam cavity, showing the 52nm slot width, air-clad waveguide coupled geometry, and input/output grating couplers. (b) Field and power distributions for the fundamental mode of the slotted photonic crystal single nanobeam cavity with Q-factor $\sim 3.5 \times 10^4$ and mode volume $\sim 0.025(\lambda/n)^3$. The plotted distributions are for a 2D slice through the center of the cavity, taken from 3D FDTD simulation.

FIG. 3 (a) Transmission spectra for the slotted photonic crystal single nanobeam cavity, revealing the cavity resonance at 1513.9nm and dielectric band edge at ~1548nm. Inset reveals IR camera images (aligned to top SEM image) spanning the cavity and output grating coupler for different wavelengths of interest: on-resonance, in the PBG, and in the dielectric band. (b) Radiation spectra for the same device in (a), with the inset revealing a close up of the resonance at 1513.9nm with a Lorenztian fit indicating an experimentally measured Q-factor, $Q_{meas} \sim 7.42 \times 10^3$.

Figures:

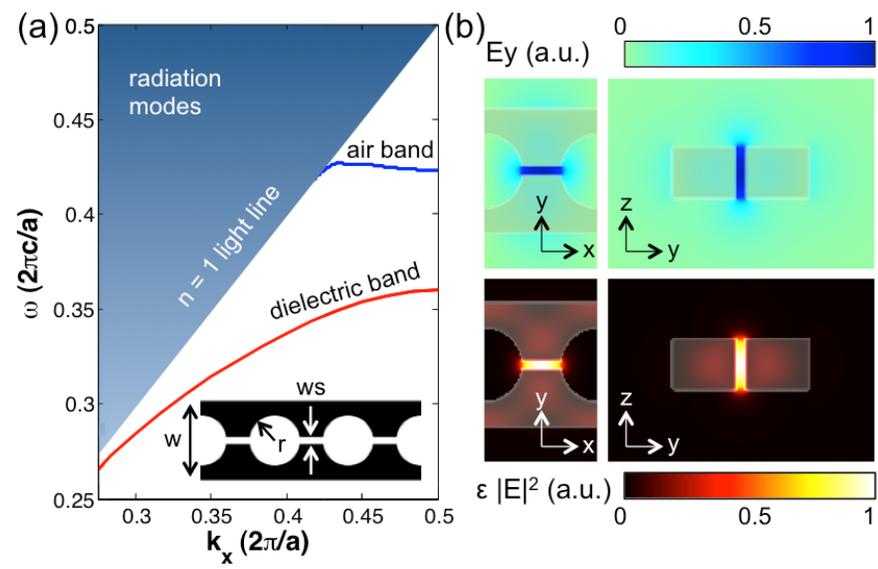

Fig. 1

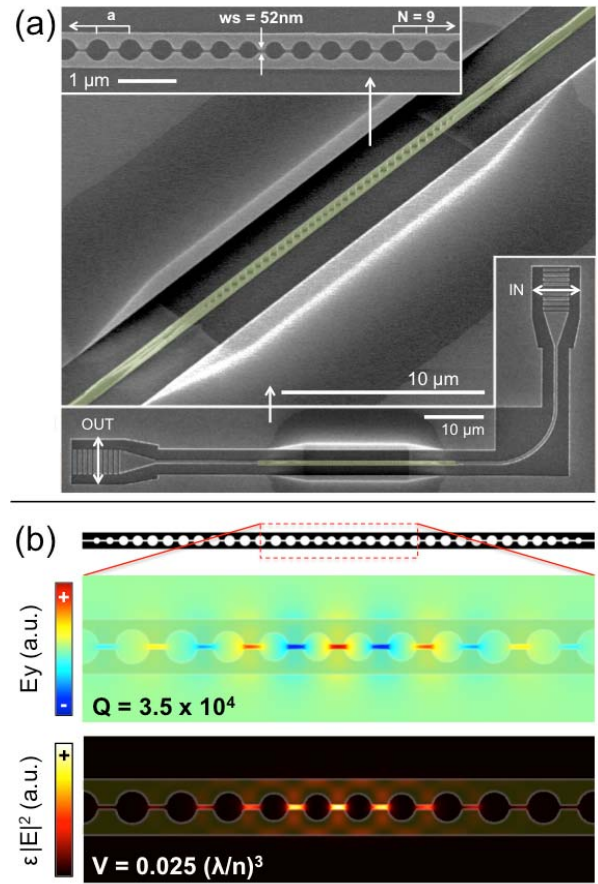

Fig. 2

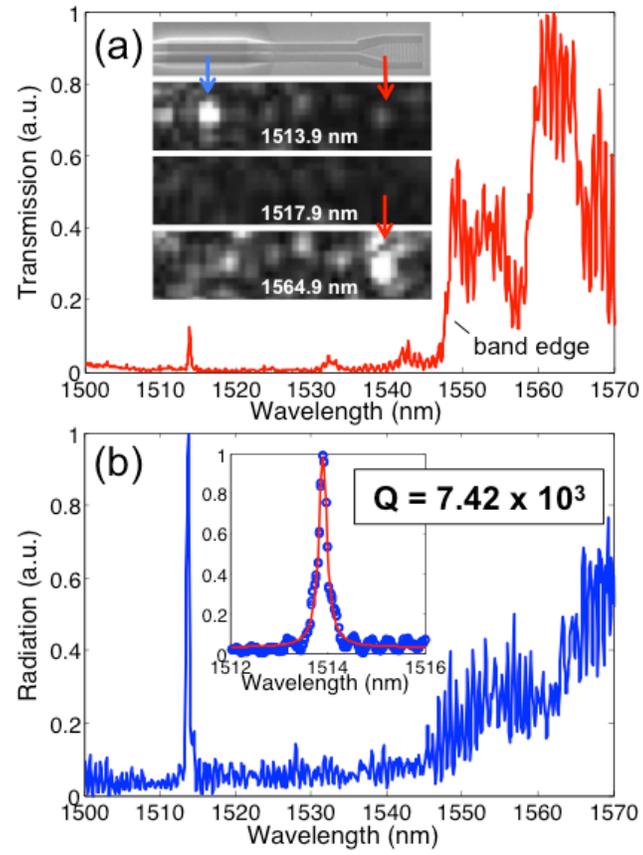

Fig. 3